\documentclass[final]{aipproc}
\layoutstyle{6x9}

\begin{document}

\title{Electromagnetic properties of baryon resonances}

\classification{14.20.Gk, 13.75.Gx, 13.60.Le } \keywords {Baryon
resonances, transition form factors}

\author{Lothar Tiator}{
address={Institut f\"ur Kernphysik, Johannes
Gutenberg-Universit\"at, D-55099 Mainz, Germany}}

\begin{abstract}
Longitudinal and transverse transition form factors for most of the
four-star nucleon resonances have been obtained from high-quality
cross section data and polarization observables measured at MAMI,
ELSA, BATES, GRAAL and CEBAF. As an application, we further show how
the transition form factors can be used to obtain empirical
transverse charge densities. Contour plots of the thus derived
densities are shown and compared for the Roper and $S_{11}$ nucleon
resonances.
\end{abstract}

\maketitle

\section{Introduction}

During the last decade, significant progress on the electromagnetic
excitation of nucleon resonances has been obtained. For pion and eta
photoproduction very precise data of unpolarized cross sections and
photon asymmetries were measured at MAMI@Mainz, ELSA@Bonn,
LEGS@Brookhaven and GRAAL@Grenoble. For electroproduction, at Mainz,
Bonn and Bates measurements for the $\Delta(1232)$ excitation were
performed at low $Q^2$, and for higher $Q^2$ up to about
$6$~GeV$^2$, at JLab a very large amount of data was collected in an
energy range up to the third nucleon resonance region.

In parallel with the ongoing experiments, several theoretical groups
developed models and analysis techniques, which were applied to the
data. The model-inde\-pen\-dent GWU/SAID
analysis~\cite{Workman:2011vb} mostly analyzed the pion
photoproduction data and improved the values of the photon couplings
over the years. Coupled channels analyses were performed by the
Giessen group~\cite{Shklyar:2006xw} and by the Bonn-Gatchina
group~\cite{Anisovich:2009zy}. Transition amplitudes were also
determined in the framework of dynamically generated resonances by
coupling to meson-baryon channels by the
J\"ulich~\cite{Doring:2009uc} and EBAC~\cite{Suzuki:2010yn} groups.

However, most successful concerning the general applicability to the
higher resonances, were the unitary isobar models of the Mainz group
(MAID model)~\cite{Maid98,MAID07,Tiator:2011pw} and of the JLab
group~\cite{Aznauryan:2011qj} who used dispersion relations as an
additional constraint to reduce the model dependence due to
incomplete experimental input.

With our unitary isobar model MAID, we analyzed all available
electroproduction data in order to determine the transition form
factors for all four-star resonances below $W=1.8$~GeV. In most
cases we could obtain both single-$Q^2$ and $Q^2$ dependent
transition form factors for the proton target. In the case of the
neutron, the parametrization of the $Q^2$ dependence had to take a
simpler form because of the much smaller world database.

Single-$Q^2$ data points for longitudinal and transverse form
factors have been obtained for transitions from the proton to the
$\Delta(1232),\,P_{11}(1440),\,S_{11}(1535),\,D_{13}(1520),$
$S_{31}(1620),\,S_{11}(1650),\,D_{15}(1675),\,F_{15}(1680),
\,D_{33}(1700)$ and $P_{13}(1720)$, which can be downloaded from the
MAID website~\cite{MAID}. Here only results for the
$\Delta(1232),\,P_{11}(1440)$, and the $S_{11}(1535)$ are shown.
Full results and details of the parameterizations are given in our
recent review article, Ref.~\cite{Tiator:2011pw}.

The main motivation for exploring the nucleon transition form
factors is to obtain a precise knowledge of the nucleon excitation
spectrum, which provides -- together with the elastic form factors
-- a complete description of the nucleon's electromagnetic
structure. This structure can be compared with QCD inspired quark
models and, in recent years, more and more also with lattice QCD
calculations.

\subsection{First resonance region}

\begin{figure}
\includegraphics[width=12cm]{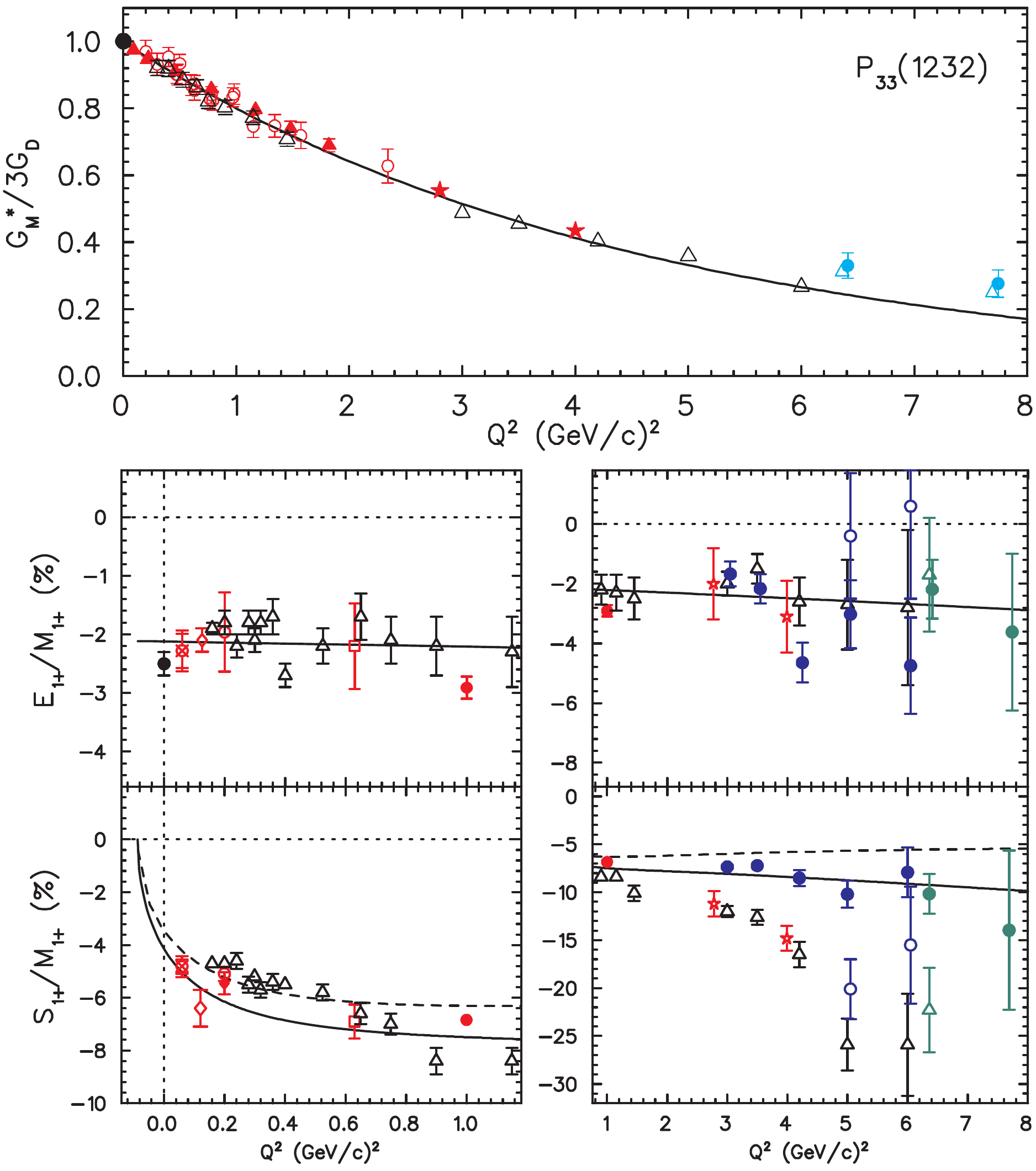}
\vspace{3mm} \caption{\label{fig:p33ffs}Electromagnetic form factors
of the $N$ to $\Delta(1232)$ transition. The upper panel shows the
magnetic form factor $G_M^*$ in the definition of Ash and the middle
and lower panels show the $E/M$ and $S/M$ ratios, respectively. The
curves are the result of our MAID2007 fit to the full dataset of
pion electroproduction. The data points are obtained in analyses at
fixed $Q^2$. The open symbols show the results of the JLab
analysis~\cite{Aznauryan:2011qj}, all others are obtained in our
MAID analysis. The two largest $Q^2$ values ($6.4,7.8$) show the
MAID and the JLab analyses of the cross sections by Villano et
al.~\cite{Villano:2009sn}. For further details, see our review,
Ref.~\cite{Tiator:2011pw}.}
\end{figure}

The $\Delta(1232)$ is the only nucleon resonance with a well-defined
Breit-Wigner resonance position, $M_R=1232$~MeV. It is an ideal
single-channel resonance, the Watson theorem applies, and the
Breit-Wigner position coincides with the K-matrix pole position. For
these reasons, the $N\to\Delta(1232)$ form factors can be determined
in an essentially model independent way. The magnetic form factor
shown in Fig.~\ref{fig:p33ffs} is very well known already from
inelastic electron scattering up to high momentum transfer,
$Q^2=10$~GeV$^2$, and can be parameterized in a surprisingly simple
form found in our previous MAID analysis,
\begin{equation}
G_M^*(Q^2)=3\,G_D(Q^2) e^{-0.21 Q^2/{\rm {GeV}}^2}\,
\end{equation}
with $G_D$ the standard dipole form factor of the nucleon.

The electric and Coulomb form factors are much smaller and are
usually given as ratios of $G_E^*$ and $G_M^*$ to the magnetic form
factor in percent. Figure~\ref{fig:p33ffs} also compares the
MAID2007 solutions (solid lines) for the ratios $R_{EM}$ and
$R_{SM}$ with other analyses. The ratio $R_{EM}$ from MAID2007 stays
always below the zero line, in agreement with the original analysis
of the data~\cite{Ung06,Fro99} and also with the dynamical model of
Sato and Lee~\cite{SL01} who concluded that $R_{EM}$ remains
negative and tends towards more negative values with increasing
$Q^2$ instead of an uprise towards unity. This indicates that the
predicted helicity conservation at the quark level is irrelevant for
the present experimental $Q^2$ range. For the ratio $R_{SM}$ our
$Q^2$ dependent fit approaches a negative constant for large $Q^2$
in good agreement with the predictions of Ji et
al.~\cite{Ji:2003fw}, Buchmann~\cite{Buchmann:2004ia} and Ramalho et
al.~\cite{Ramalho:2008dp}, who use a relation between the ratio
$R_{SM}$ and the ratio of the electric and magnetic neutron form
factors. For $Q^2>1~{\rm {GeV}}^2$ our Maid2007 single-$Q^2$
analysis for $R_{SM}$ disagrees with the JLab analysis of Aznauryan
et al.~\cite{Azn09}. Whereas our analysis stays almost constant, the
JLab analysis suggests a much larger negative slope. By repeating
our data analysis at $Q^2=5$ and $6$~GeV$^2$ in different energy
ranges, we found a strong dependence of the fit on the energy
interval used. Our results of 2007 were obtained with the full
energy range of the measured data, $W=1110-1390$~MeV.

Figure~\ref{fig:p33ffs} also shows an analysis (blue open circles)
in the energy range of $W=1200-1260$~MeV, much closer to the
resonance position. If we choose the energy interval even closer to
resonance, $W=1220-1240~$MeV, the errors increase further by a
factor of 2 and the $E/M$ ratio becomes large and positive, while
the $S/M$ ratio remains the same. We conclude that the analysis in
this $Q^2$ range strongly depends on the energy interval and the
parametrization of the background used in the analysis and requires
further studies.

\subsection{Results for the second resonance region}

\begin{figure}
\includegraphics[width=14.0cm]{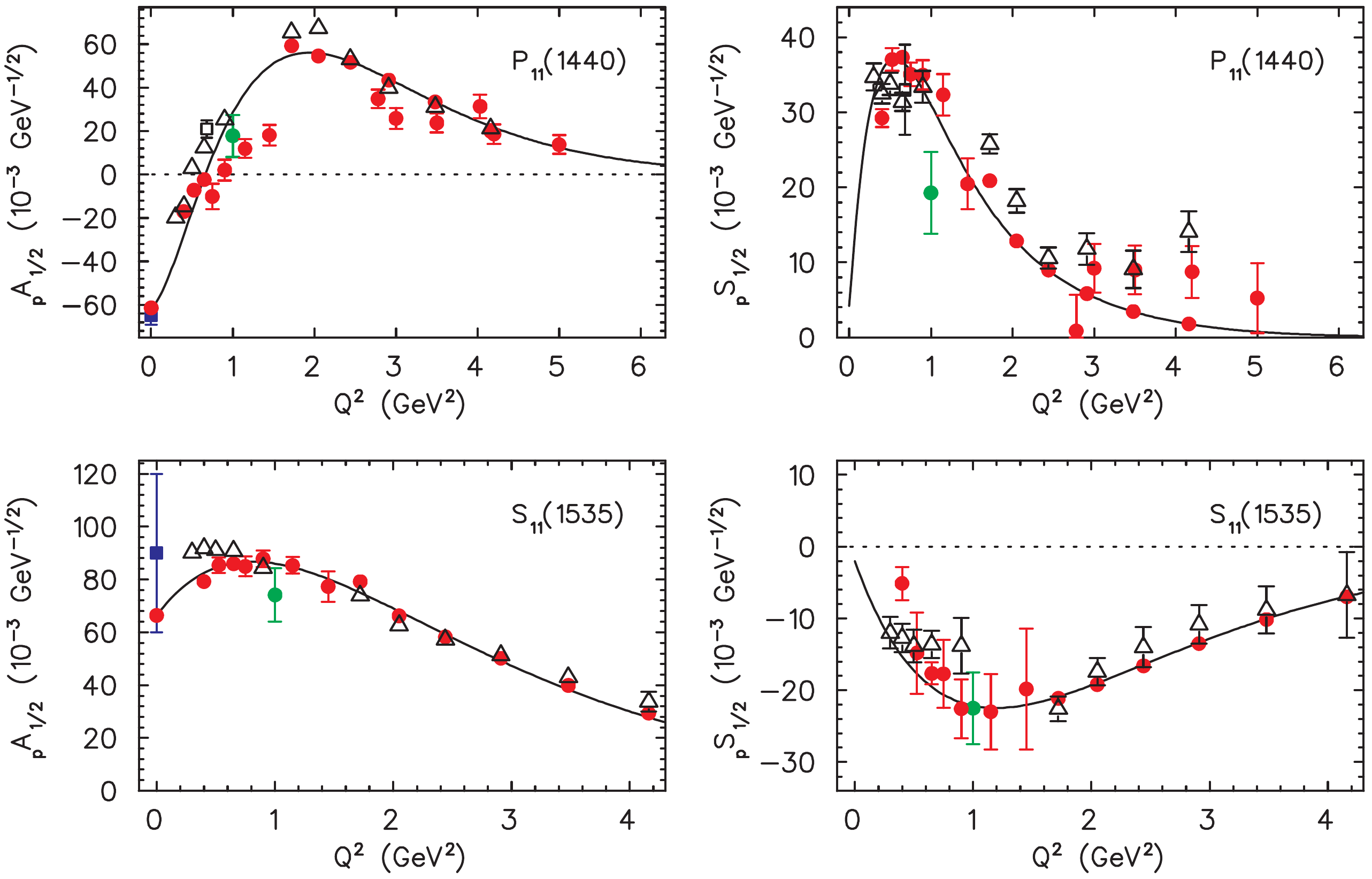}
\caption{\label{fig:p11s11ffs} Transverse ($A_{1/2}$) and
longitudinal ($S_{1/2}$) electromagnetic form factors of the $p$ to
$P_{11}(1440)$ (upper panel) and  $p$ to $S_{11}(1535)$ (lower
panel) transitions. The curves are the result of our MAID2007 fit to
the full dataset of pion electroproduction. The data points are
obtained in analyses at fixed $Q^2$. The open symbols show the
results of the JLab analysis~\cite{Azn09,Villano:2009sn}, all others
are obtained in our MAID analysis. For further details, see
Ref.~\cite{Tiator:2011pw}.}
\end{figure}

The helicity amplitudes for the Roper resonance $P_{11}(1440)$ are
displayed in the upper panel of Fig.~\ref{fig:p11s11ffs}. Our latest
$Q^2$ dependent solution (solid lines) is in reasonable agreement
with the single-$Q^2$ analysis (red circles). The figure shows a
zero crossing of the transverse helicity amplitude $A_{1/2}(Q^2)$ at
$Q^2\approx 0.7$~GeV$^2$ and a maximum at the relatively large
momentum transfer $Q^2\approx 2.0$~GeV$^2$. The longitudinal Roper
excitation $S_{1/2}(Q^2)$ rises to large values around $Q^2\approx
0.5$~GeV$^2$ and produces one of the strongest longitudinal
amplitudes that we find in our analyses.

The lower panel in Fig.~\ref{fig:p11s11ffs} displays the results for
the $S_{11}(1535)$ resonance. The red single-$Q^2$ data points show
our results of 2007, which are in good agreement with our $Q^2$
dependent analysis (solid lines). The black triangles are the 2009
results of Ref.~\cite{Azn09}. The blue data point at $Q^2=0$
represents the PDG average over several $\gamma,\pi$ and
$\gamma,\eta$ analyses. While we find values around 65 in all MAID
analyses, the JLab analysis obtains values around 90 for
$\gamma,\pi$ and 110 for $\gamma,\eta$~\cite{Aznauryan:2011qj}. Also
the SAID and Bonn-Gatchina groups extract values around 100, but in
a very recent analysis, based on pion photoproduction, Shrestha and
Manley report, however, also a small coupling of
$59(3)$~\cite{Shrestha:2012ep}.

From the $Q^2$ dependent parametrization of the transition form
factors we can calculate transverse charge transition densities, as
viewed from a light front moving towards the
baryon~\cite{Tiator2009}. For  that we first have to transform the
helicity form factors $A_{1/2},S_{1/2}$ to the Dirac-like form
factors $F_1^{NN^\ast}$ and $F_2^{NN^\ast}$, which results in simple
linear relations. The transverse densities are then obtained by a
2-dim Fourier-Bessel transformation. The densities relating to $F_1$
appear as fully symmetrical monopole patterns, while the $F_2$ form
factors transform to additional dipole patterns (further details in
Ref.~\cite{Tiator:2011pw}).

\begin{figure}[htbp]
\includegraphics[width=6.9cm]{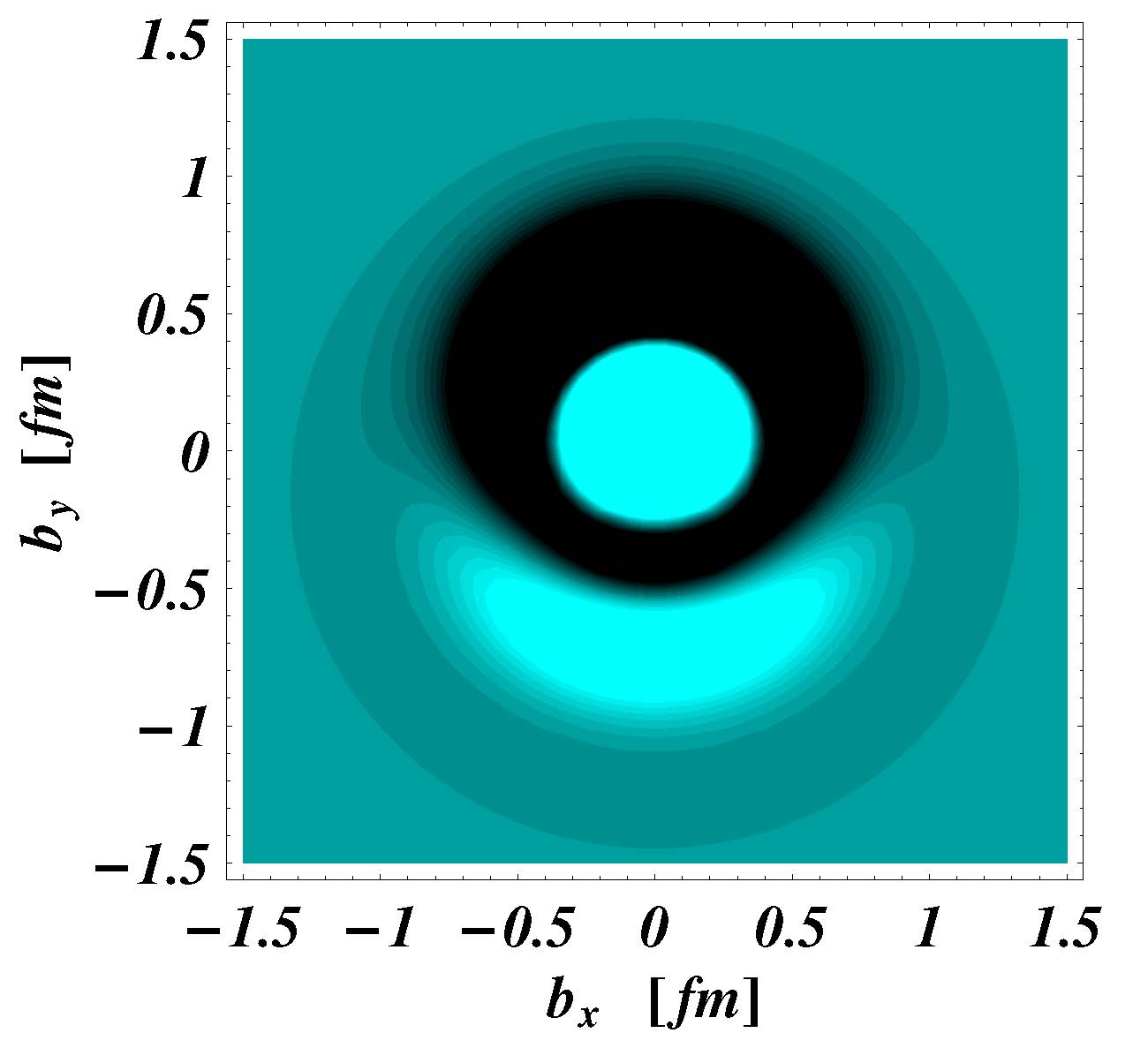}
\hspace{0.1cm}
\includegraphics[width=6.9cm]{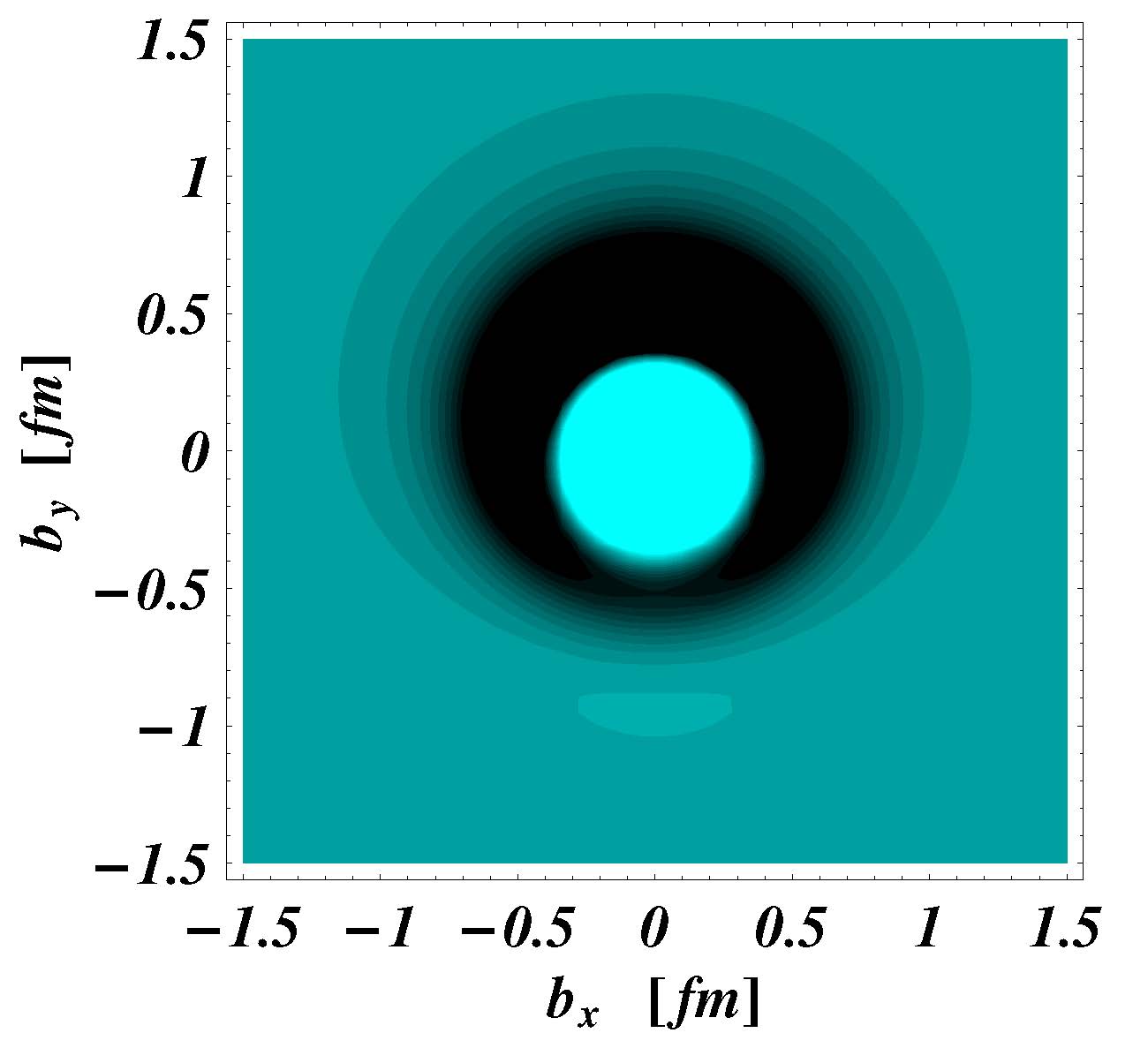}
\vspace{3mm} \caption{\label{fig:p11s11dens1p} Comparison of the
polarized quark transverse transition charge densities corresponding
to the e.m. transitions $p \to P_{11}(1440)$
($\frac{1}{2}^+\rightarrow\frac{1}{2}^+$) (left panel) and $p \to
S_{11}(1535)$ ($\frac{1}{2}^+\rightarrow\frac{1}{2}^-$) (right
panel). The light (dark) regions correspond to positive up quark
(negative down quark) densities. }
\end{figure}

As an example, in Fig.~\ref{fig:p11s11dens1p} we show the polarized
quark transverse charge densities from the proton to the Roper and
to the $S_{11}(1535)$ resonances. Comparing these results, we find
that the dipole contribution of the up and down quarks to the
polarized densities is much less pronounced for the $S_{11}$ due to
the much smaller $F_2^{NN^*}/F_1^{NN^*}$ form factor ratio.

\begin{theacknowledgments}
We want to thank the Deutsche Forschungsgemeinschaft for the support
by the Collaborative Research Center 1044.
\end{theacknowledgments}

\bibliographystyle{aipproc}

\end{document}